\documentclass[12pt]{iopart}

\usepackage{graphicx}
\begin{document}

\title[Quantum correlations for Ising- Heisenberg Model on a symmetrical diamond chain]{ Investigation of the Quantum Correlations for a S=1/2 Ising- Heisenberg Model on a Symmetrical Diamond Chain}

\author{E. Faizi, H. Eftekhari}

\address{Physics Department, Azarbaijan shahid madani university}
\ead{efaizi@azaruniv.edu}
\ead{h.eftekhari@azaruniv.edu}
\begin{abstract}
 We consider the quantum correlations for a  S=1/2 Ising- Heisenberg model of a symmetrical diamond chain. Firstly, we compare concurrence, quantum discord and 1- norm geometric quantum discord of an ideal diamond chain ($J_m=0$) in the absence of magnetic field. The results show no simple ordering relations between these quantum correlations, so that quantum discord may be smaller or larger than the 1-norm geometric quantum discord, which this observation contradict the previous result that provided by F. M. Paula \cite{F. M. Paula}. Symmetrical behavior of quantum correlation versus ferromagnetic and anti- ferromagnetic coupling constant J is considerable. The effect of external magnetic field H and temperature- dependence is also considered. Furthermore, we study quantum discord and geometric measure of quantum discord with the effect of next nearest neighbor interaction between nodal Ising sites for a generalized diamond chain (${J_m}\neq0$), and we observe coexistence of phases with different values of magnetic field for quantum correlations.  Moreover, entanglement sudden death occurs while quantum discord, 1- norm geometric quantum discord and geometric quantum discord are immune from sudden death.

\end{abstract}

\maketitle

\section{Introduction}
The quantum correlations has some advantages in comparison with classical ones which cause a rapid progress of quantum information
and communication devices. But, the specifying of quantum correlations in a given system is not easy task.
At first, the quantum entanglement was considered as a good measure for quantifying quantum correlations \cite{Werner, Hill, Peres, Amico}. Later, the quantum discord (QD) was introduced
as a more convenient measure of quantum correlations \cite{W, L, W. H}. Particularly, there is  states with zero entanglement which can exhibit non- zero QD.
 However, the calculation of QD is very difficult because of optimization problem. In general, despite of intensive study of quantum QD only very special cases
 have been solved analytically \cite{F, M, S. L, Ali, M. Ali}. To avoid this hardness, Dakic et al. \cite{B} proposed a geometric measure of quantum discord (GMQD) which
 measures the quantum correlations of a quantum system base on the minimal Hilbert- Schmidt distance between the given state and a state with zero discord
, and they derived an definitive formula for specifying the GMQD for any two qubit state in 2012. Furthermore, it has been indicated to exhibit operational
importance in certain quantum communication protocols (see, refs. \cite{Y. O, M. Gu}). Nevertheless, as recently pointed out \cite{Hu, T. Tufarelli, M. Piani}, GMQD as introduced in ref. \cite{B}
can not be consider as an appropriate measure for quantumness of correlations, because it may increase by local operations on the unmeasured subsystem. Specifically, it has  been revealed by Piani in ref. \cite{M. Piani} that the introduction of a local ancilary state on the unmeasured subsystem may alters the geometric discord by a factor given by the purity of the ancilla. Essentially, the root of this problem is the absence of contractivity of the Hilbert- Schmidt norm under quantum channels which are trace- preserving. Notably, this problem can be solved if Schatten 1- norm (trace norm) is applied as a
distance measure \cite{B. Aaronson, Aaronson, T. Nakano}.\\
In recent years, many efforts have been paid to determine the quantum correlations properties of condensed matter systems,
which are the convenient candidate to utilize for quantum communication and quantum information. Therefore, it is very significant to investigate quantum correlation of solid state systems such as spin chains \cite{K. M}. The Heisenberg  spin chain as a simplest quantum systems has been studied in many subjects of quantum information and computation.\\
In the last decade, many diamond chain structures have been studied. Pereira et al. \cite{Pereira, M. S. S} studied the magnetization effect in kinetically frustrated diamond chains. Recently, Lisnii \cite{Lisnii} investigated a distorted diamond Ising- Hubbard chain, and that the model also exhibits the geometrical frustration. Thermodynamics of the Ising- Heisenberg model of a diamond- shape chain was debated in the refs. \cite{Canova, Rojas, J. S . Valverde, O. Rojas, B. M. Lisnii} greatly.

In this work, we investigate the quantum correlations for a S=1/2 Ising- Heisenberg model of a symmetrical diamond chain. The effect of magnetic field and next nearest neighbor interaction nodal Ising sites is discussed. In Sec. 2 we briefly review the definition of QD, GMQD and 1- norm geometric quantum discord (1- norm GQD) respectively. Also, in Sec. 3 we present  S=1/2 Ising- Heisenberg model on a generalized symmetrical diamond. The quantum correlation properties of an ideal diamond chain ${J_m=0}$ are discussed in Sec. 4. The next section contains similar contents with the incorporation of ${J_m}$ interaction. Moreover, some results are given in Sec. 6.
\section{quantum discord, entanglement, GMQD and Schatten 1-norm GQD}
\subsection{quantum discord }
By having a quantum state $\rho$ in a composite Hilbert space ${H=H_A\otimes{H_B}}$, the amount of total correlation is determined by quantum
mutual information \cite{Groisman}.
\begin{eqnarray} I(\rho)=S(\rho_A)+S(\rho_B)-S(\rho),
\end{eqnarray}
\\
In which ${S(\rho)=-Tr(\rho\log_2{\rho})}$ is the Von Neumann entropy and ${\rho_{A(B)}=Tr_{B(A)}\rho}$ is the reduced matrix
by tracing out the subsystem B(A). If we consider the part A as the apparatus, QD is determined as follows \cite{L, H. Ollivier}.
\begin{eqnarray} D(\rho)=I(\rho_A)-C_A(\rho),
\end{eqnarray}
which is the difference between total amount of correlation ${I(\rho|\{E_K\})}$ and the classical correlation. Classical correlation is defined by \cite{L, H. Ollivier, N. Li}
\begin{eqnarray} C(\rho)=\max_{E_K}{I(\rho|\{E_K\})},
\end{eqnarray}
Where ${I(\rho|\{E_K\})}$ is a variant of quantum mutual information of a given measurement basis $\{E_K\}$ on the subsystem A as follows:
 \begin{eqnarray} {I(\rho|\{E_K\})=S(\rho_B)-\Sigma_K{P_K}S(\rho_{B|K})},
\end{eqnarray}
${\rho_{B|K}=Tr_A[(E_K\otimes{1})\rho]/p_K}$ is the post measurement state of subsystem B after acquiring the outcome K on part A with the probability ${p_K=Tr[(E_K\otimes{1})\rho]}$. $\{E_K\}$ correspond to a set of one dimensional projectors on ${H_A}$ and 1 is the identity operator.
\subsection{Entanglement}
In order to investigation the dynamic of two-qubit entanglement, we use Wootter's concurrence \cite{W. K. Wootters}. For two qubits system, the concurrence is calculated from the density
matrix $\rho$ for qubits A and B:
\begin{eqnarray} C({\rho})=max{\{0,\sqrt{\lambda_1}-\sqrt{\lambda_2}-\sqrt{\lambda_3}-\sqrt{\lambda_4}}\},
\end{eqnarray}
Where the quantities  ${\lambda_i}$ are the eigenvalues in decreasing order of the matrix ${\xi}$:
\begin{eqnarray} \xi=\rho(\sigma_y\otimes\sigma_y)\rho^*(\sigma_y\otimes\sigma_y),
\end{eqnarray}
Where $\rho^*$ indicates the complex conjugation of $\rho$ in the standard basis $|00\rangle,|01\rangle,|10\rangle,|11\rangle$ and $\sigma_y$ is the Pauli matrix.

\subsection{GMQD}
Dakic et. al. proposed the geometric measure of quantum discord of the state ${\rho}$ defined by \cite{B}
\begin{eqnarray} D_G({\rho})=\min_{\chi}{{\|\rho-\chi\|}^2},
\end{eqnarray}
where the minimum is taken over the set of zero discord states (i.e., ${D(\chi)=0}$) and ${{\|\rho-\chi\|}^2}=Tr(\rho-\chi)^2$ is the square norm
in the Hilbert- Schmidt space.
For any two qubit state
\begin{eqnarray} \rho^{AB}={\frac{1}{4}}[I\otimes{I}+\sum_{i=1}^3(x_i\sigma_i\otimes{I}+y_iI\otimes{\sigma_i})+\sum_{i,j=1}^3(r_{ij}\sigma_i\otimes{\sigma_i})],
\end{eqnarray}
its GMQD is given by
\begin{eqnarray} D_G({\rho})={\frac{1}{4}}{(\|X\|^2+\|R\|^2-K_{max})},
\end{eqnarray}
where $\sigma_i$ are the pauli spin matrices, ${X=(x_1,x_2,x_3)^T}$, R is the matrix elements ${r_{ij}}$, and ${k_{max}}$ is the maximal eigenvalue of the matrix
${K=XX^T+RR^T}$.
\subsection{Schatten 1-norm GQD}
 We consider a bipartite system AB in a Hilbert space  ${H={H_A}\bigotimes{H_B}}$. The system is determined by quantum states characterized by density operators ${\rho\in{B(H)}}$, where B(H) is the set of bound, positive-semidefinite operators acting on H with ${Tr[\rho]=1}$. The 1-norm GQD between A and B is defined through the trace distance between ${\rho}$ and the closest classical- quantum state ${\rho_c}$ \cite{F. M. Paula, B. Aaronson, Aaronson, T. Nakano},  reading
\begin{eqnarray} D_G({\rho})=\min_{\Omega_{0}}{{\|\rho-\rho_c\|}_1},
\end{eqnarray}
where ${\|X\|_1=Tr{[\sqrt{X^\dag{X}}]}}$ is the 1-norm (trace norm) and ${\Omega_0}$   is the set of classical-quantum states.

 In the certain case of two- qubit Bell diagonal states, whose density operator possess the form
 \begin{eqnarray} \rho={\frac{1}{4}}{[{I}\otimes{I}+\vec{c}.({\vec{\sigma}}\otimes{\vec{\sigma}})]},
\end{eqnarray}
Where I is the identity matrix, ${\vec{c}=(c_1,c_2,c_3)}$ is a three- dimensional vector such that ${-1\leq{c_i}\leq1}$ and ${\vec{\sigma}=(\sigma_1,\sigma_2,\sigma_3)}$ is a vector composed by Pauli matrices.

, 1- norm GQD can be analytically computed, yielding \cite{F. M. Paula}
\begin{eqnarray}D_G({\rho})=int[|c_1|,|c_2|,|c_3|],
\end{eqnarray}
where ${int[|c_1|,|c_2|,|c_3|]}$ is the intermediate result among the absolute values of the correlation functions ${c_1,c_2}$ and ${c_3}$.

\section{S=1/2 Ising- Heisenberg model for a generalized symmetrical diamond}
We use the S=1/2 Ising- Heisenberg model for a generalized symmetrical diamond chain, which hamiltonian for this model can be written as follows \cite{Ananikian}:
 \begin{eqnarray} H=\sum_{k=1}^N{H_k}&=&\sum_{k=1}^N[J_2S_{k1}S_{k2}+J(\mu_k^z+\mu_{k+1}^z)(S_{k1}^z+S_{k2}^z)\nonumber\\ &+&J_m\mu_k^z\mu_{k+1}^z-H(S_{k1}^z+S_{k2}^z+\frac{\mu_k^z+\mu_{k+1}^z}{2})],
\end{eqnarray}
Where $H_k$ denotes the Hamiltonian of k- th cluster, $S_k=(S_k^x,S_k^y,S_k^z)$ indicates the Heisenberg spin 1/2 operator, $\mu_k$ is the Ising spin. $J,J_2,J_m>0$ corresponds to the anti-ferromagnetic coupling constants.

The reduced density (by tracing out of two spin of four spin of the cluster) matrix $\rho$ for k- th cluster \cite{Ananikian}:
\begin{eqnarray}
\rho={\frac{1}{Z}} \left(
\begin{array}{cccccccccccccccc}
u && 0 && 0 && 0\\
0 && w && y && 0\\
0 && y && w && 0\\
0 && 0 && 0 && v\\
\end{array}
\right).
\end{eqnarray}
Where
\begin{eqnarray} u=2e^{\frac{4H+J_m-J_2}{4T}}+e^{-\frac{-2H+J_m-4J+J_2}{4T}}+e^{-\frac{-6H+J_m+4J+J_2}{4T}};\nonumber\\
v=e^{-\frac{6H+J_m+J_2+4J}{4T}}(2e^{\frac{H+J_m-4J+2J}{2T}}+e^{\frac{H+2J}{T}}+1);\nonumber\\
w={\frac{1}{2}}(e^{\frac{J_2}{T}}+1)e^{-\frac{2H+J_m+J_2}{4T}}(2e^{\frac{H+J_m}{2T}}+e^{\frac{H}{T}}+1);\nonumber\\
y={-\frac{1}{2}}(e^{\frac{J_2}{T}}-1)e^{-\frac{2H+J_m+J_2}{4T}}(2e^{\frac{H+J_m}{2T}}+e^{\frac{H}{T}}+1),
\end{eqnarray}
And Z is the partition function:
 \begin{eqnarray}Z=u+v+2w,
\end{eqnarray}

The entropy of ${\rho}$ can be easily obtained:
 \begin{eqnarray}S(\rho)=-\Sigma_{i=1}^4{\lambda_i\log_2\lambda_i},
\end{eqnarray}
where ${\lambda_i}$ are eigenvalues of the density matrix ${\rho}$.
 The minimization of conditional entropy can be acquired as refs. \cite{M. Ali, Shunlong}. It is clear that any von Neumann measurement for subsystem B can be written as ${B_i=V\Pi_iV^\dag}$, where ${\Pi_i={|i\rangle}{\langle{i}|}}$ and ${\{|i\rangle\}}$ is the standard basis ${\{|0\rangle, |1\rangle\}}$.
 ${V\in{SU(2)}}$, is a unitary operator. Any ${V\in{SU(2)}}$ can be written as ${V=tI+iy.\sigma}$, with ${t, y_1, y_2, y_3\in{R}}$ and ${t^2+y_1^2+y_2^2+y_3^2=1}$.  The final result is as follow:
 \begin{eqnarray}
\min_{B_i}S(\rho|\{B_i\})=-{\frac{1-\theta}{2}}\log_2{\frac{1-\theta}{2}}-{\frac{1+\theta}{2}}\log_2{\frac{1+\theta}{2}},
\end{eqnarray}
with ${\theta={\frac{1}{Z}}\max(|u-w|, |y|)}$. By knowing ${\theta}$ in eq. (19) the QD can be obtained easily by numerical calculation. u,w,y are functions of parameters $T, J, J_2, J_m,H$, so QD will obviously depend on $T, J, J_2, J_m,H$.
The concurrence $C(\rho)$ of such an X- state density matrix has the following form \cite{Hill, W. K. Wootters}:
 \begin{eqnarray}
C(\rho)={\frac{2}{Z}}{\max(|y|-\sqrt{uv},0)}
\end{eqnarray}

\section{Ideal diamond chain}
In this section we investigate quantum correlations of a dimeric part of an ideal diamond chain $(J_m=0)$. At first, we study the behavior of
quantum correlations at H=0. By putting H=0 in the element of density matrix $\rho$, QD easily calculate from equations which we present in sec 5. It is noticeable that in this case density matrix has a Bell diagonal shape with the bellow coefficient:
 \begin{eqnarray}
c_1=\frac{2y}{Z}, c_2=c_1, c_3=1-4{\frac{w}{Z}}
\end{eqnarray}
Therefore we can also calculate 1- norm GQD by using eq. (12).\\
In figure (1) quantum correlations are plotted as a function of temperature T and coupling constant J. We can see that, QD and concurrence decrease with growth of temperature and behave similarly to the increase of absolute value of J. Compared with variation of QD and concurrence, the variation of 1- norm GQD has a similar behavior as shown in figure 1(b). One of the noticeable difference between QD and 1- norm GQD is that by increasing coupling constant (J), QD decreases asymptotically while 1- norm GQD become revival for $|J|>{1}$ and low temperature. It is worthwhile to mention that, the phenomenon of sudden death for entanglement occurs, while QD and 1- norm decrease asymptotically. In this sense we can say that, QD and 1- norm GQD are more robust than the concurrence to large J and high temperature. In figure (2) the quantum correlations are compared versus temperature T and different values of J. The phenomenon of entanglement sudden death can be seen from figure (2) easily. We can see from figure (2) that QD and 1- norm GQD is always existent while the entanglement will vanish in some regions. When the temperature reaches some point, the entanglement will disappear. The higher the temperature is, the smaller the quantum correlations are. In other word, we can say that high temperature can diminish the quantum correlations. Also, we can see that the quantum correlations have different order as a function of temperature T. We should mention that their order do not preserved for different coupling constant J. When we fix J=0, 1- norm GQD is always larger than concurrence and quantum discord. However, for $J=1$ QD initially is larger than 1- norm GQD and concurrence for $0<T\leq{0.5}$, but for the range of $0.5<{T}\leq2$, 1- norm GQD becoming larger than QD and concurrence. Again, for $J=2$, 1- norm GQD is larger than QD and concurrence. Therefore, there is no definitive ordering relations between these quantum correlations. The observations are not in agreement with the information that provided by F. M. Paula \cite{F. M. Paula} which claim that 1- norm GQD is larger than QD. Figure (3) shows the relation between quantum correlations versus J and for different values of temperature T. All of these quantum correlations at T=0 have maximum value and by increasing temperature their amount decrease. Quantum correlations are symmetric versus ferromagnetic and anti- ferromagnetic coupling constant J. It is noticeable that, when the temperature takes a zero value, 1- norm GQD has a sudden transition from a value of 1 to about 0.26 (at J=1). By increasing temperature this transition become lesser and eventually disappear. However, for QD and concurrence no transition take place.

\begin{figure}
\includegraphics[width=2in]{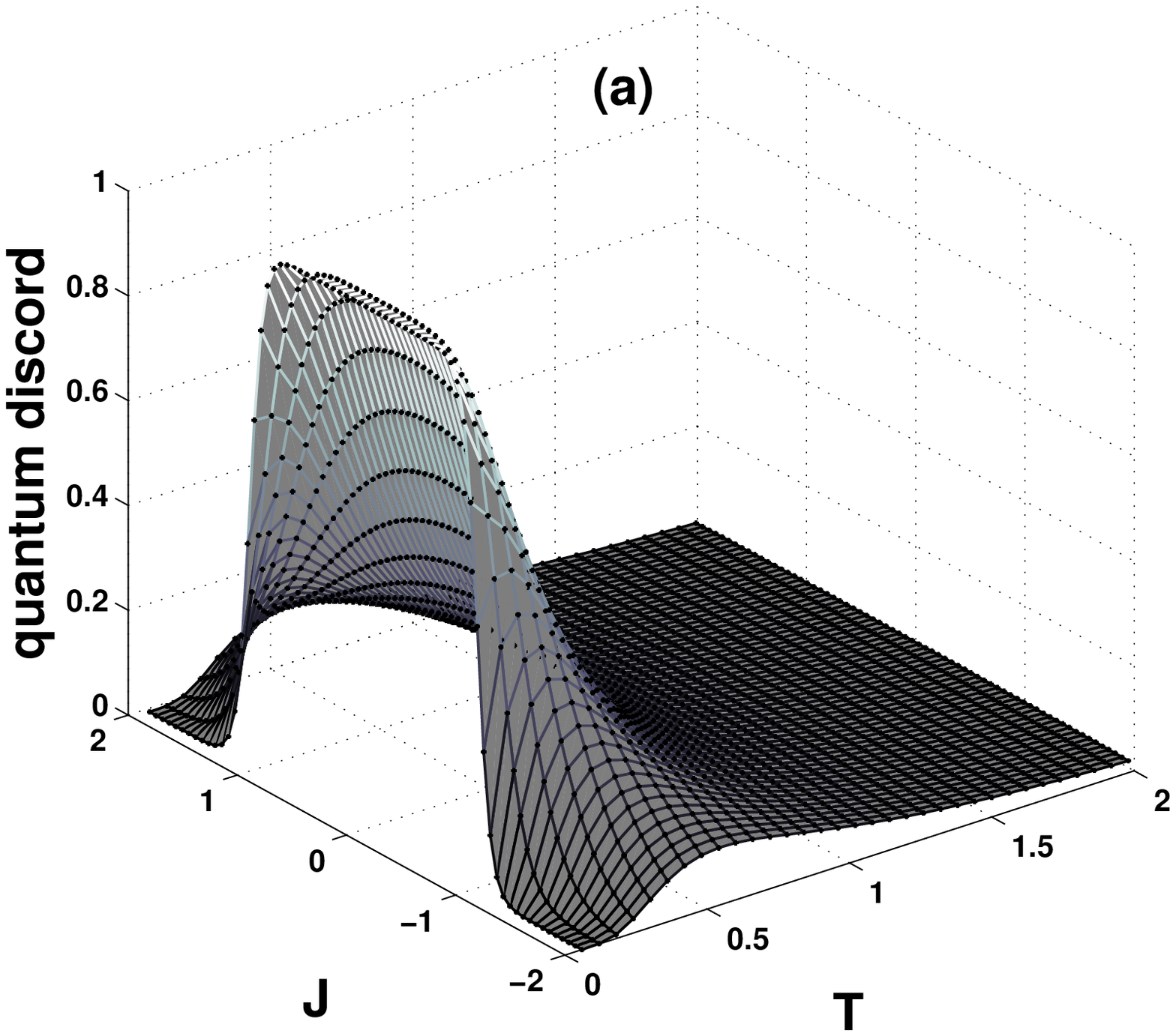}
\includegraphics[width=2in]{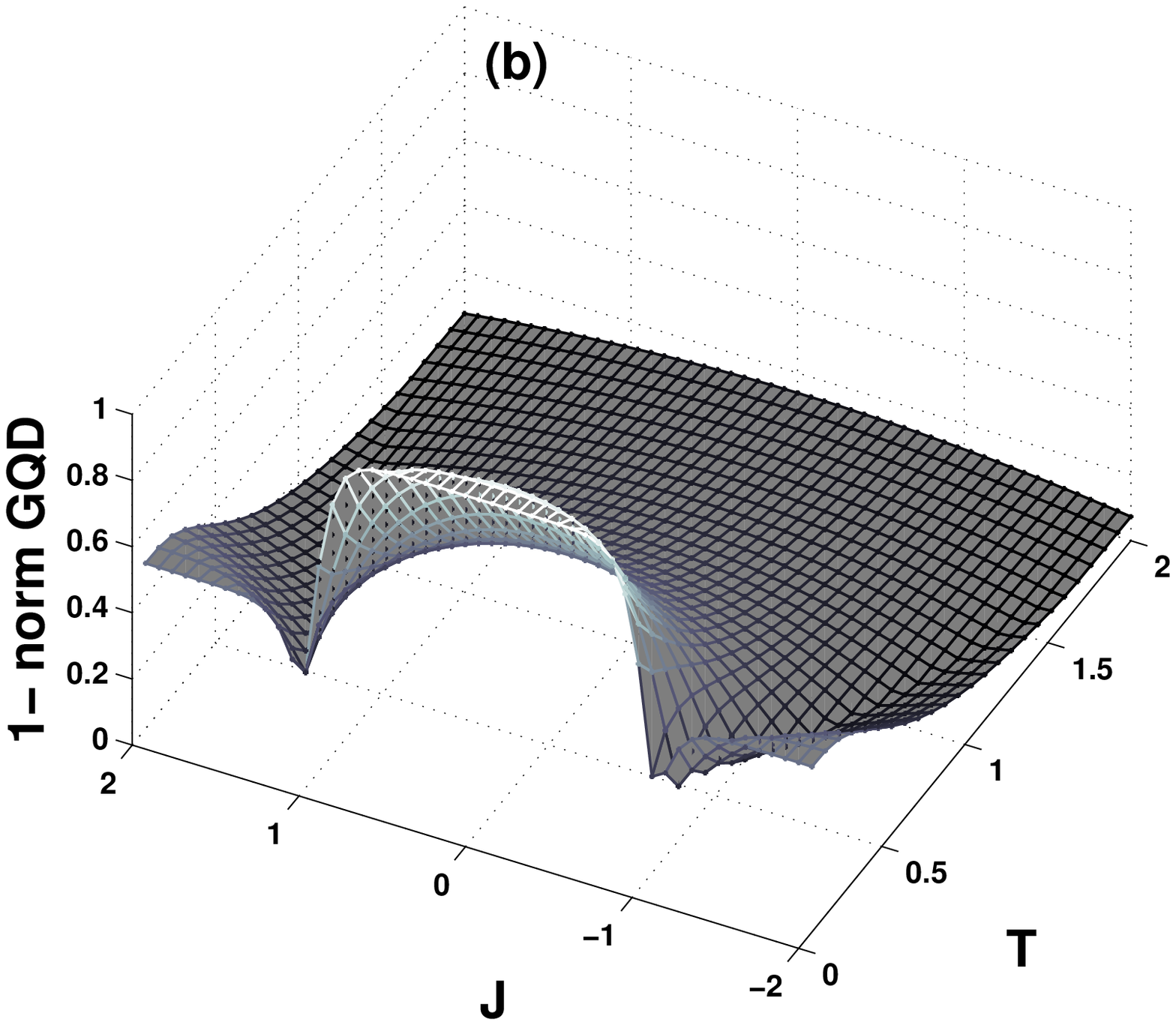}
\includegraphics[width=2in]{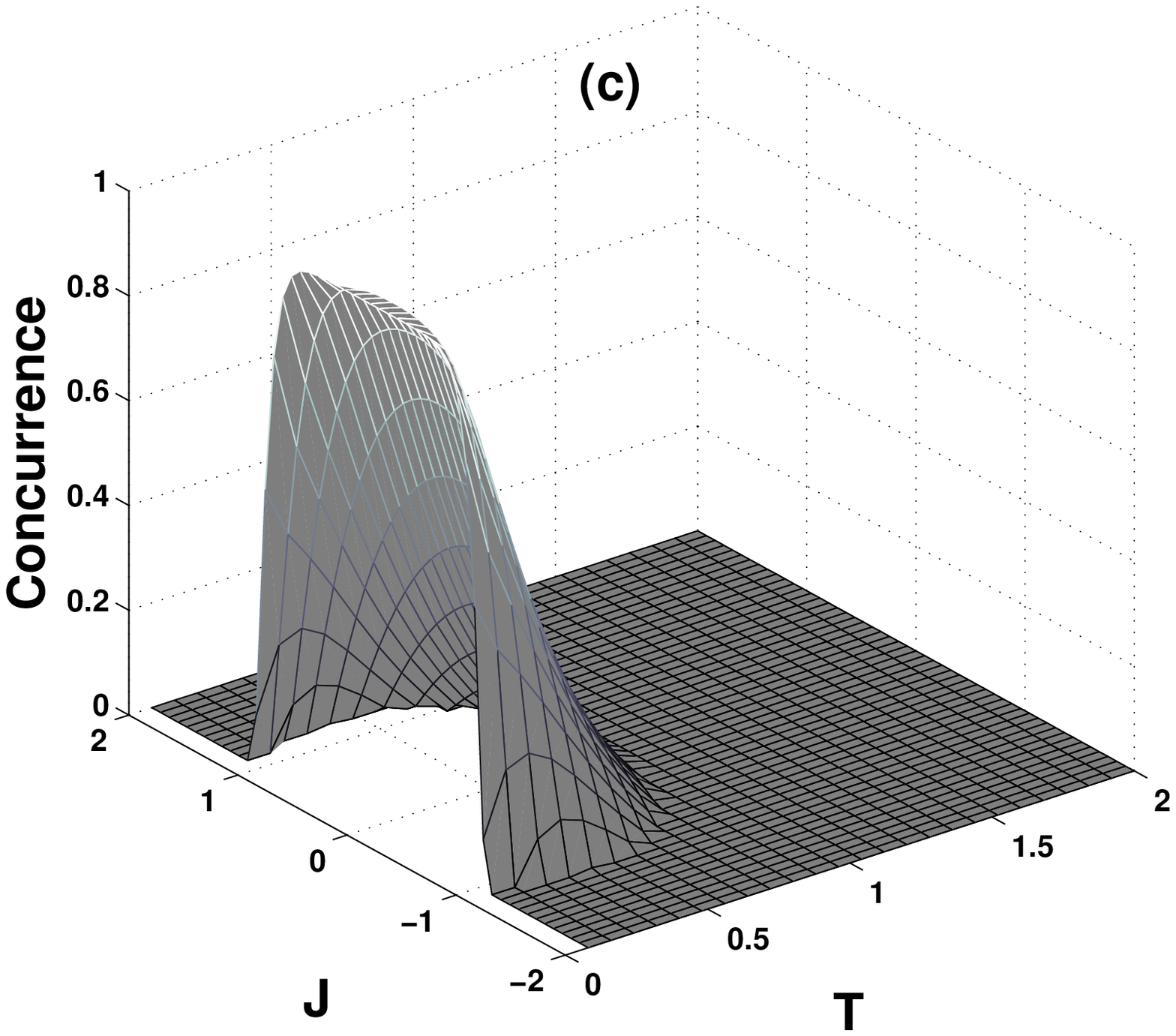}

\caption{(a) Quantum correlations versus temperature T and coupling constant J for an Ideal diamond chain ($J_m=0$) and H=0. (a) Quantum discord (b) 1- norm GQD  (c) Concurrence.}
 \label{fig1}
\end{figure}

\begin{figure}
\includegraphics[width=2in]{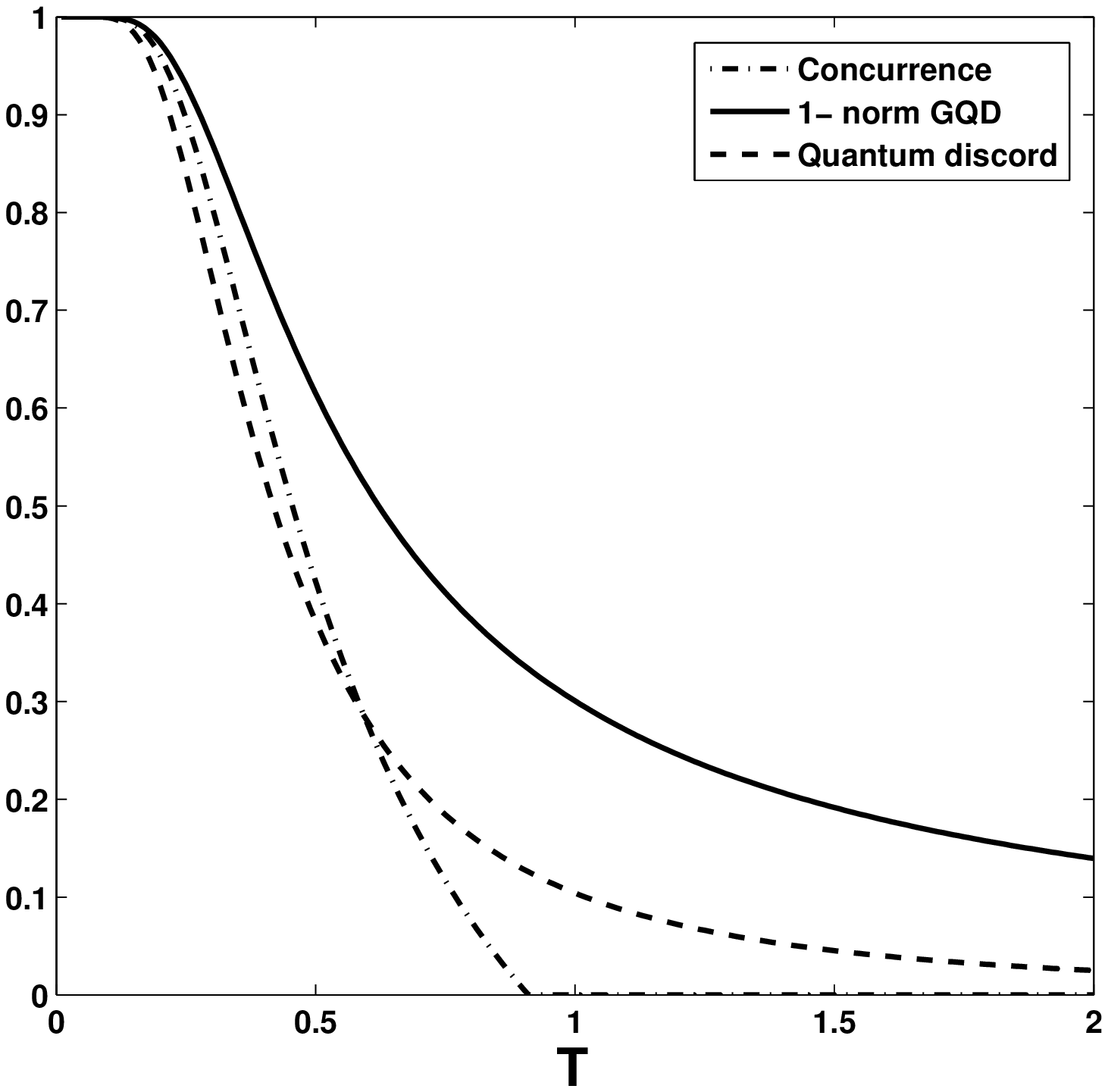}
\includegraphics[width=2in]{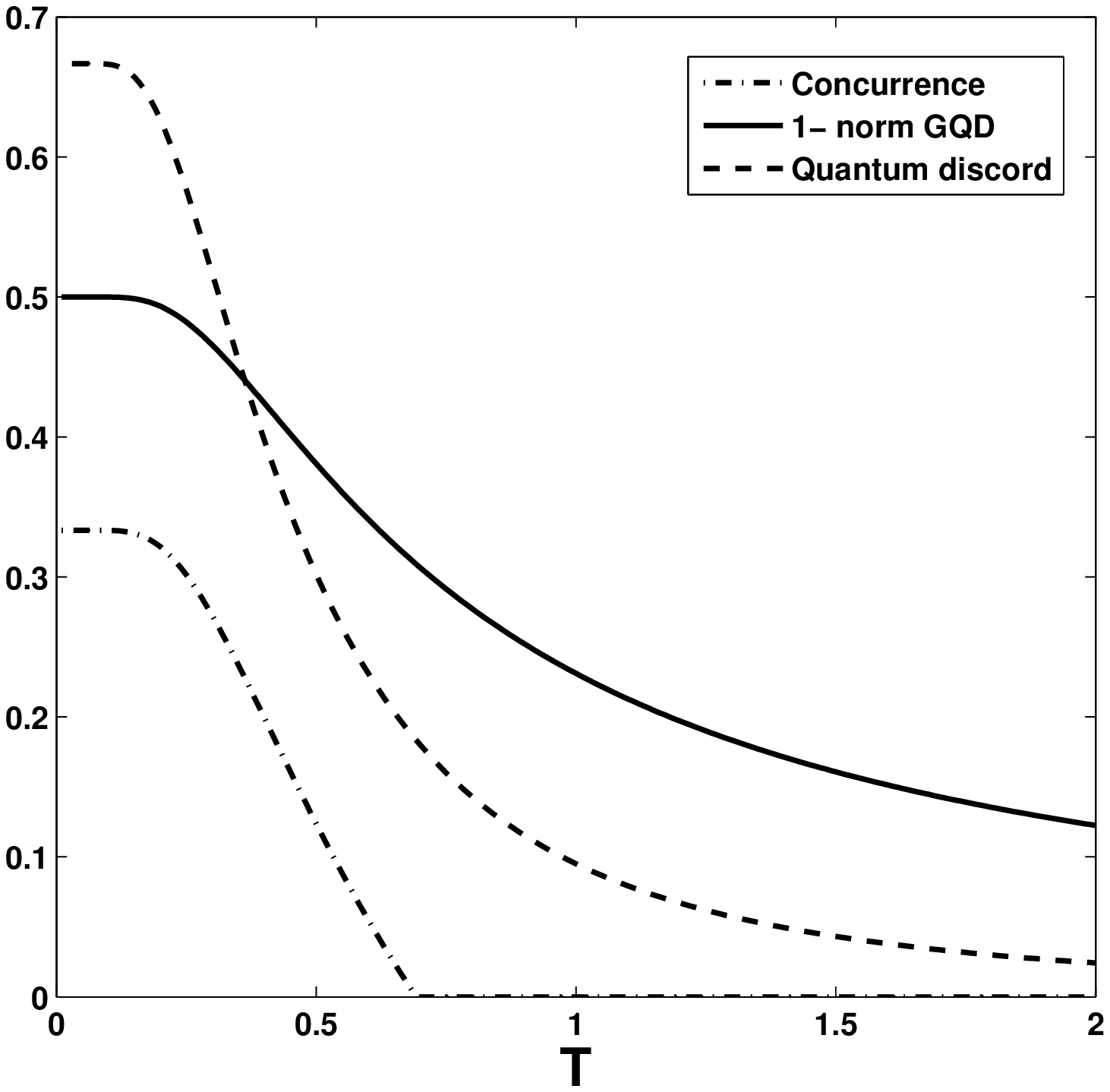}
\includegraphics[width=2in]{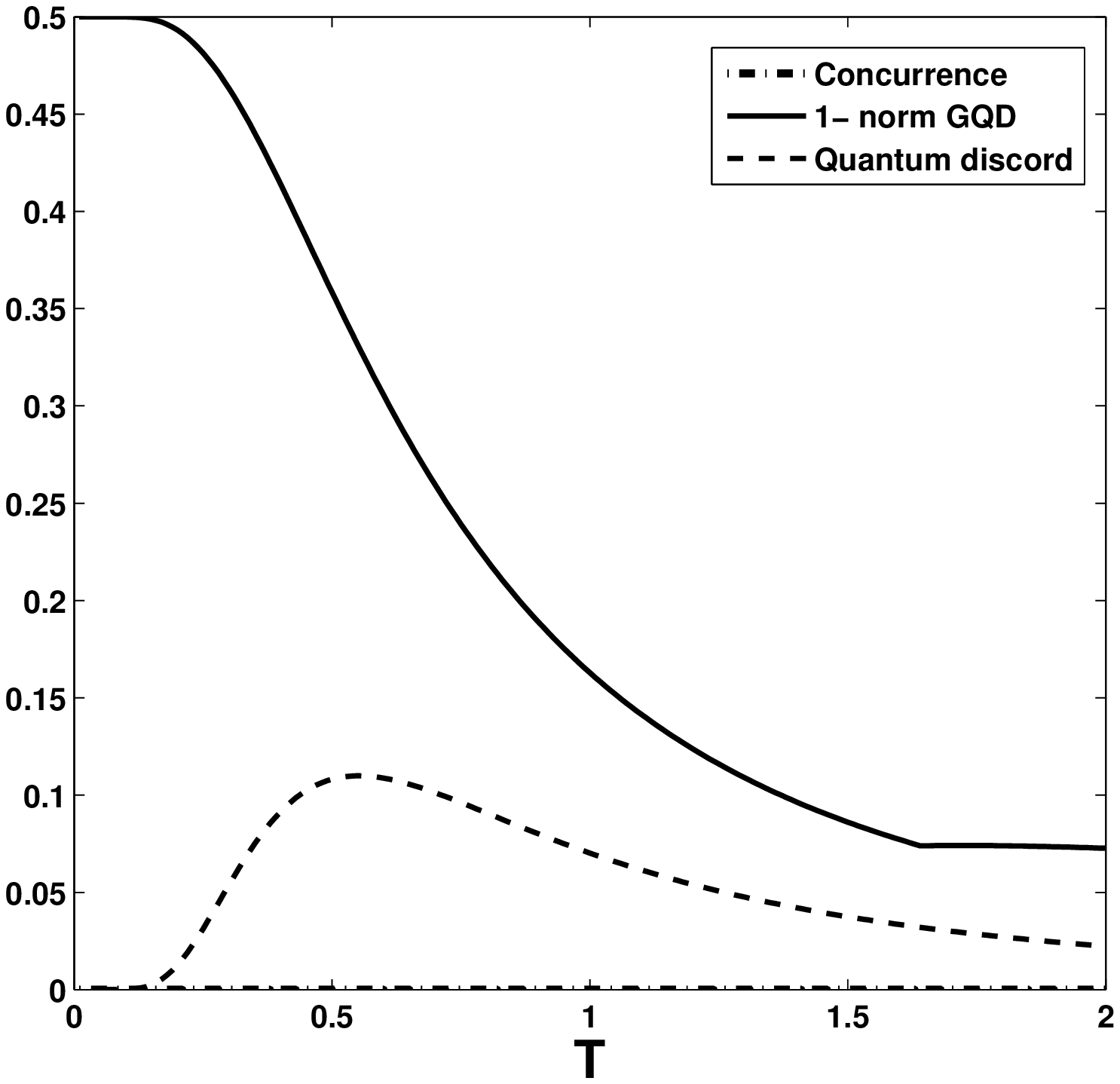}

\caption{(a) Comparison of QD, 1- norm GQD and concurrence versus temperature T for $J_2=1, J_m=0, H=0, J=0$; (b)  $J_2=1, J_m=0, H=0, J=1$ (c) $J_2=1, J_m=0, H=0, J=2$.}
 \label{fig2}
\end{figure}
\begin{figure}
\includegraphics[width=2in]{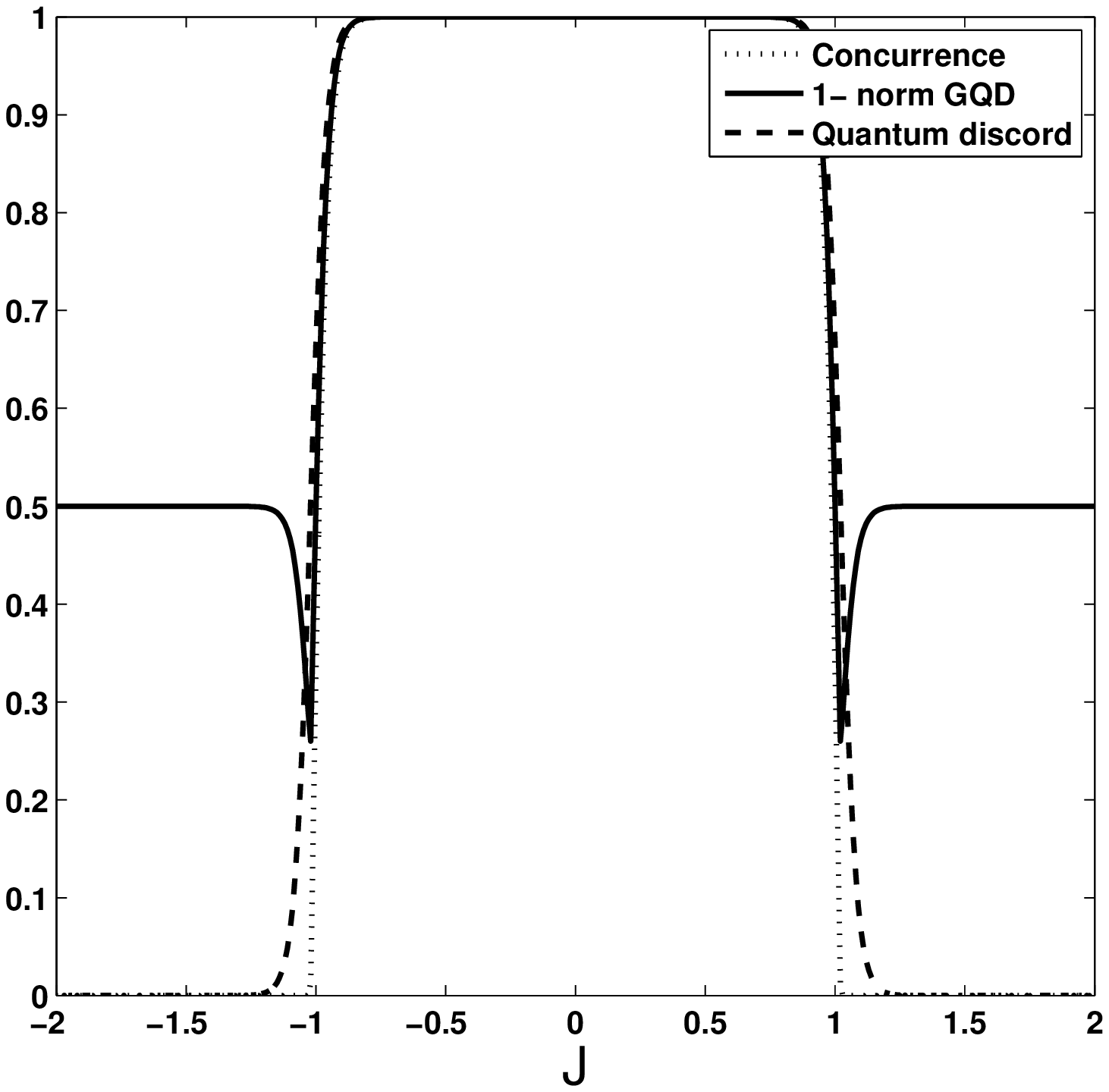}
\includegraphics[width=2in]{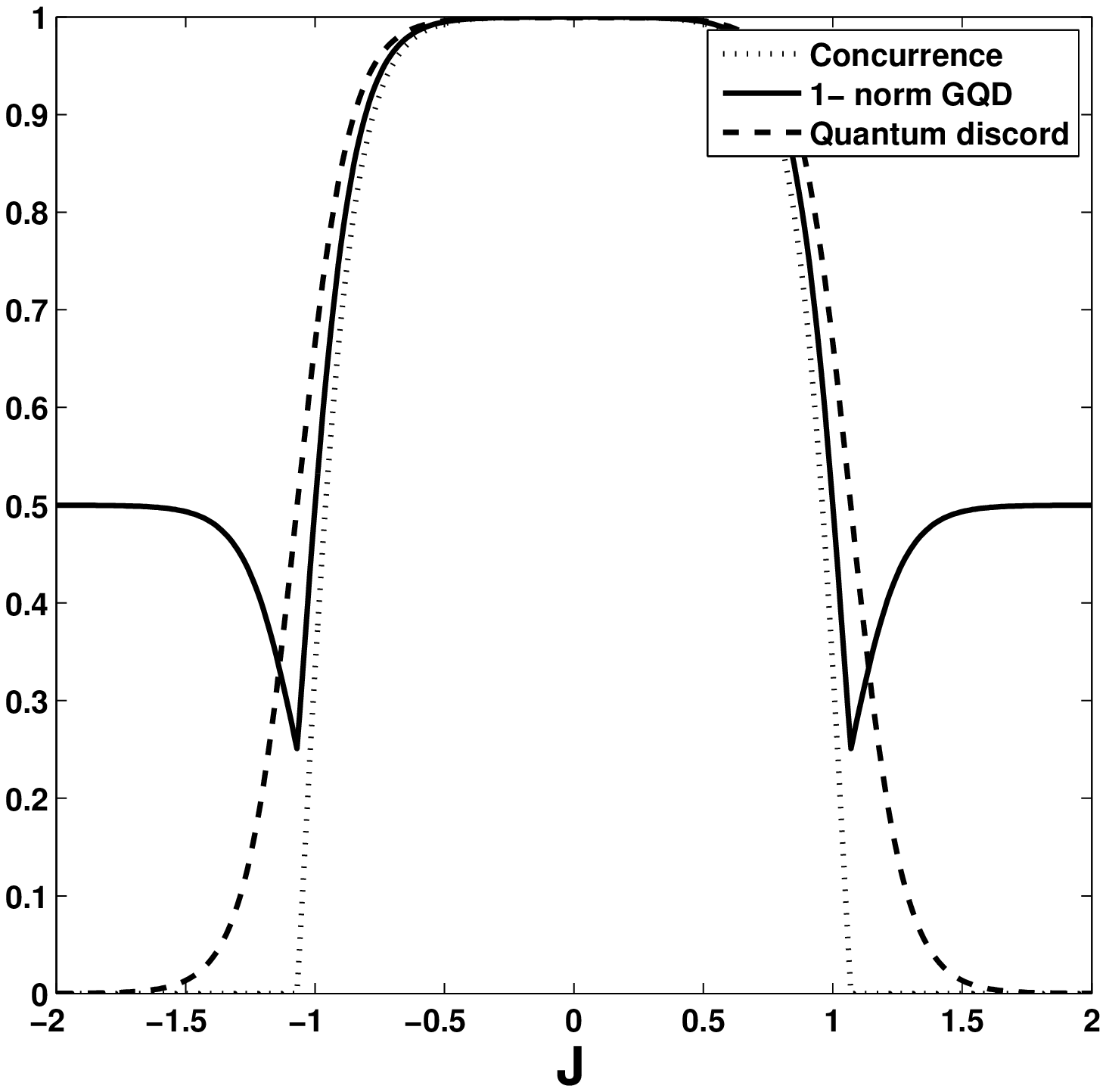}
\includegraphics[width=2in]{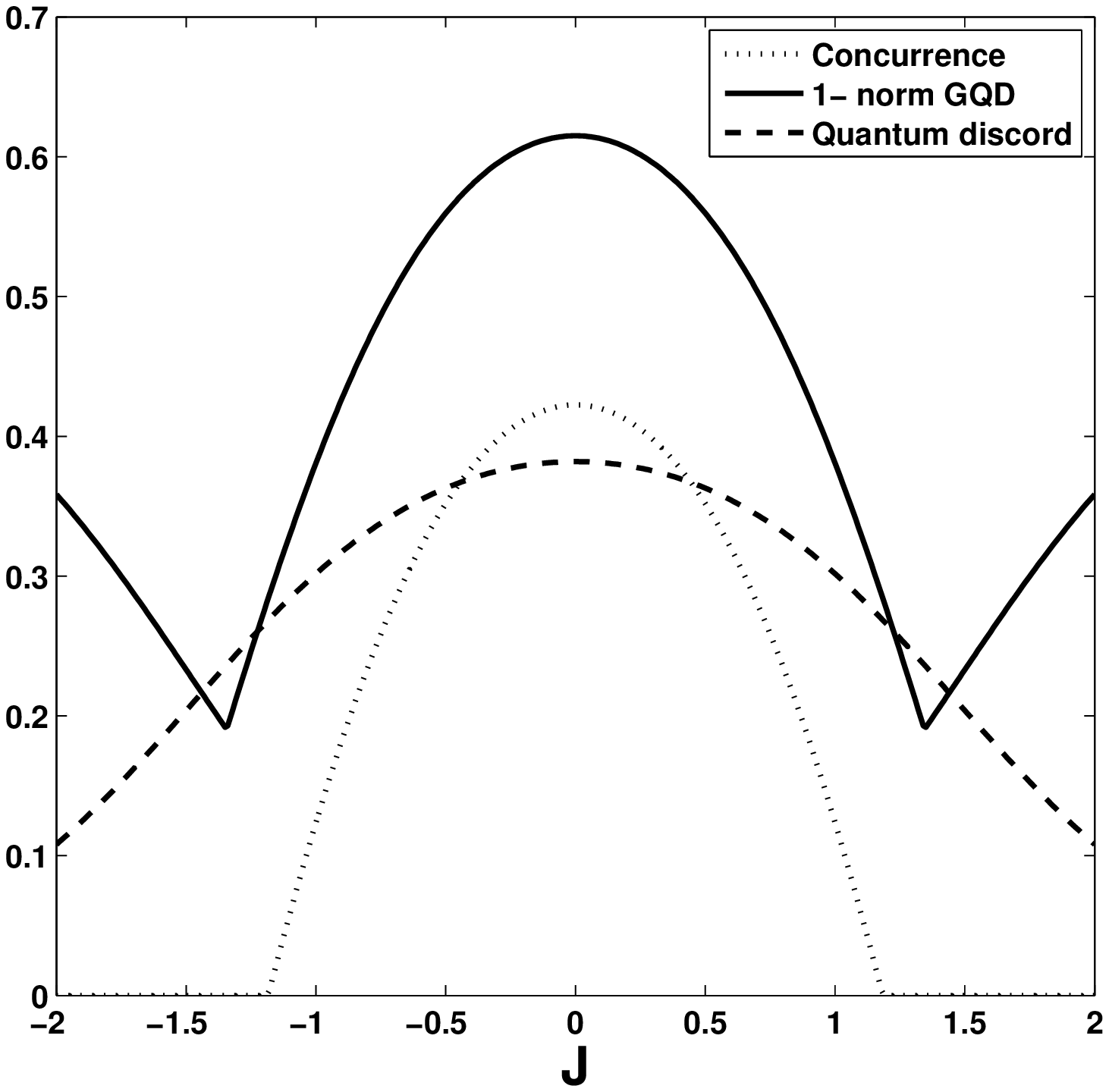}
\caption{Comparison of concurrence, quantum discord and 1- norm GQD versus J for different values of temperature, $J_2=1$, $J_m=0$ and H=0 (a) T=0, (b) T=0.1, (c) T=0.5.}
 \label{fig3}
\end{figure}

Our further investigation is the effects of the magnetic field H on quantum correlations. The dependency curve of concurrence to magnetic field at zero temperature has a dip at H=0 with $C(\rho)=1/3$ for $J-J_2=0$. Furthermore, magnetic entanglement has a higher value than that at zero magnetic field in the case $J-J_2=0$. There is no dip if $J-J_2<0$ (figure 4(a)) and in this case curve $C(\rho)$ start from $C(\rho)=1$. When Ising type interaction is stronger than the Heisenberg one $(J-J_2>0)$, one does not find a magnetic entanglement. Figure 4(b) shows the dependence of QD to magnetic field H. It is noticeable that, the behavior of QD at $J-J_2<0$ indicate no change in comparison with concurrence. However, it is not hold for $J-J_2=0$. we can see a peak at H=0, the amount of QD is equal to concurrence at this point but in other points QD is lesser than concurrence. Figure 4(c) indicate  the behavior of GMQD respect to magnetic field H. The only difference of GMQD with QD is that the amount of GMQD is very less in comparison by QD, but the general pattern is preserved. Concurrence becomes zero when $J-J_2\leq0$ at the values of magnetic field H (corresponding to saturation field $H_s^\pm$, that is when the non-entangled state becomes the ground state). Ananikian et al in \cite{Ananikian} find the value of $H_s^\pm$ ($H_s^+=J+J_2$ and $H_s^-=-J-J_2$ for $H>0$ and $H<0$ respectively). It is remarkable that, this phenomenon occurs for QD and GMQD at the same values of H (see figure (4)).

\begin{figure}
\includegraphics[width=2.8in]{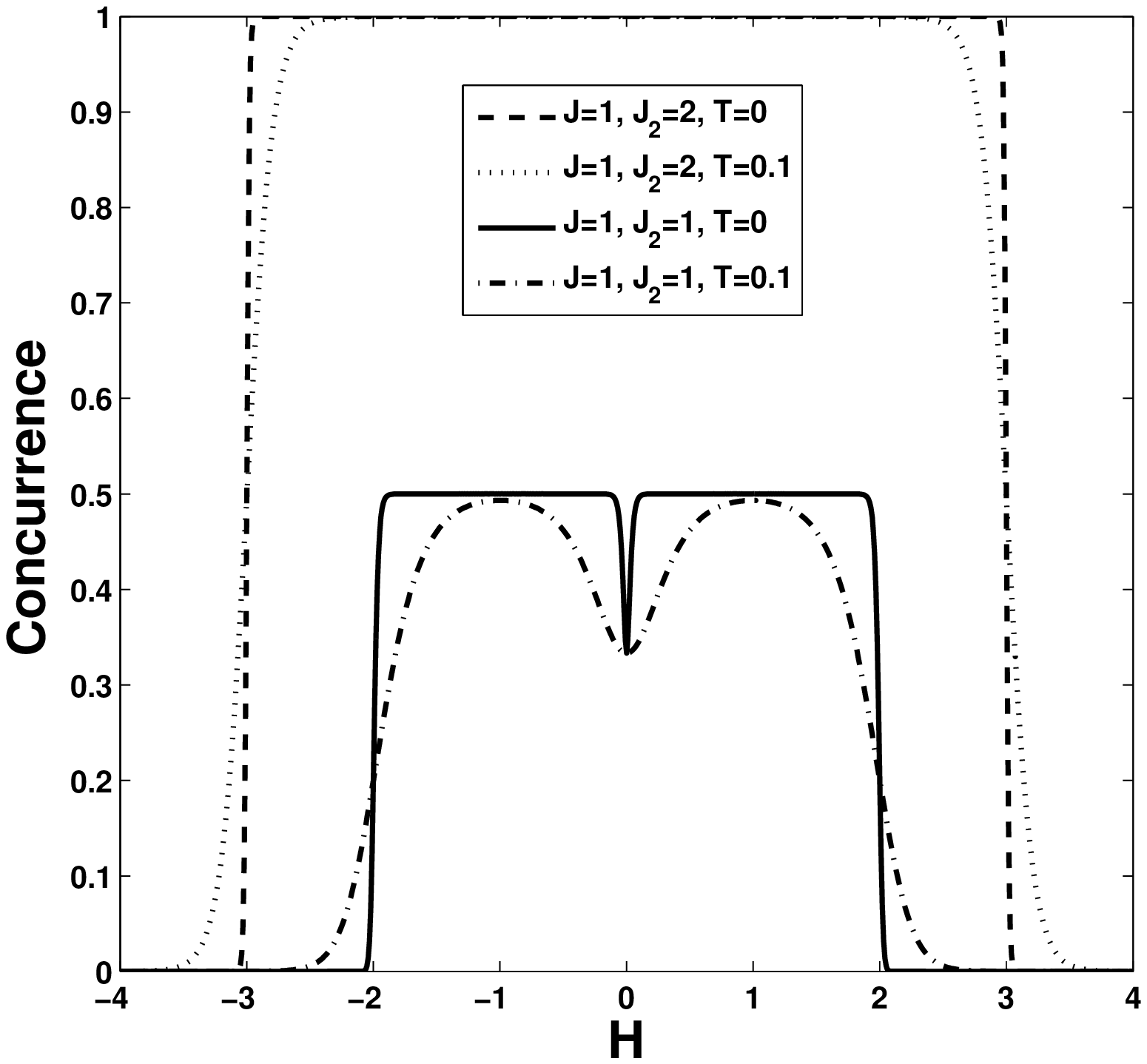}
\includegraphics[width=2.8in]{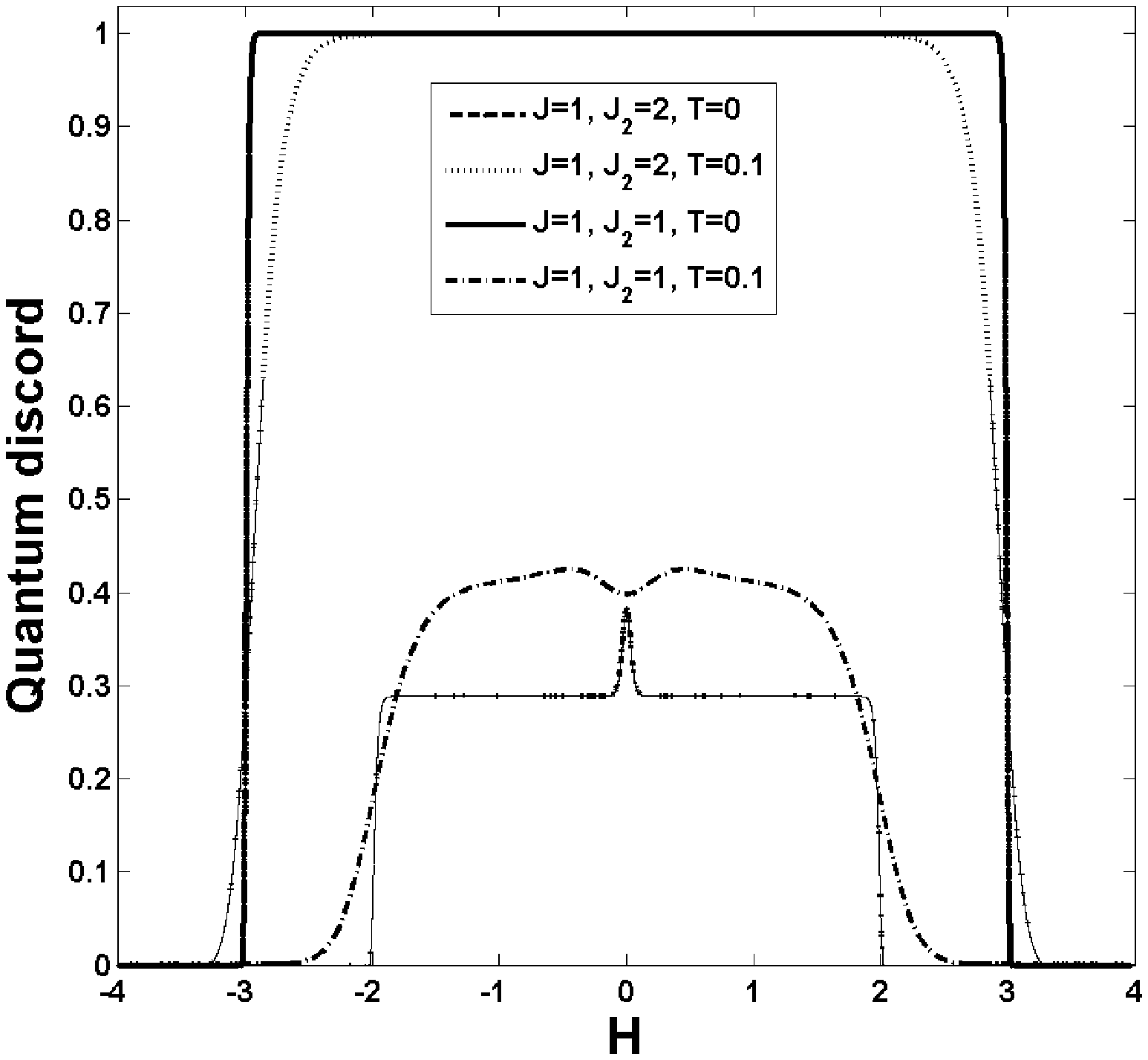}
\includegraphics[width=2.8in]{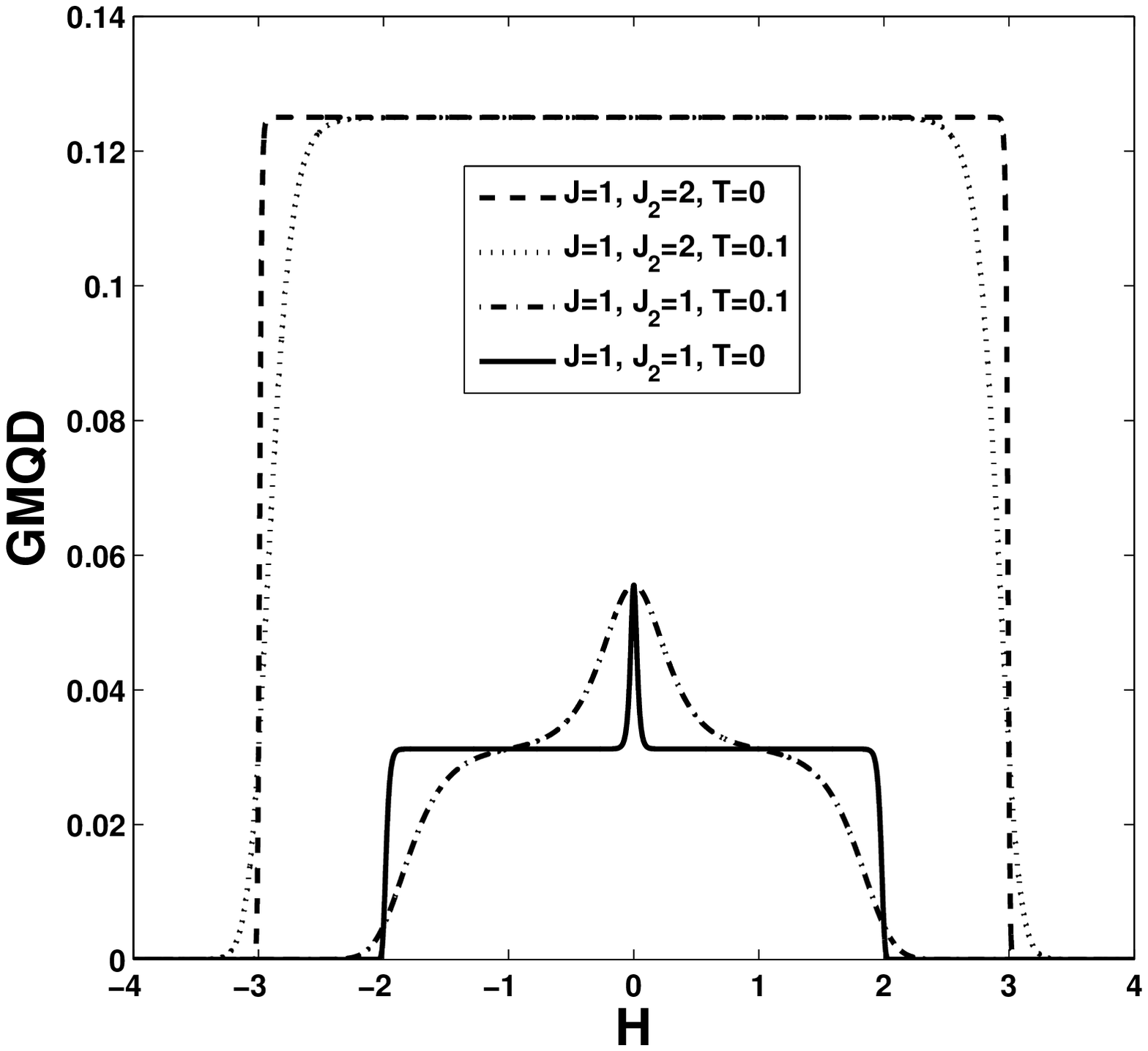}
\caption{Quantum correlations versus magnetic field for different values of temperature, $J_2$ and J. (a) Concurrence (b) Quantum discord (c) GMQD }
 \label{fig4}
\end{figure}

\section{Generalized diamond chain}
 Here we will investigate the effects of the next nearest neighbor interaction $(J_m)$ between the Ising spins of the cluster. We will discuss the regime $J-J_2>0$ introducing effects of the magnetic field H. When $J_m\leq2(J-J_2)$, one does not find magnetic entanglement in the system. Nevertheless, when $J_m>2(J-J_2)$, quantum correlations start with the maximum value at T=0 therefore we obtain magnetic entanglement. We can introduce the critical values of magnetic field $H_c^+$ and $H_c^-$, corresponding to vanishing of magnetic entanglement. We should mention that, Ananikian in ref. \cite{Ananikian} find these values ($H_c^+=2J_2-2J+J_m$ and $H_c^-=-H_c^+$). We can see from figure (5) that for all quantum correlations which we consider in this parer, critical values of magnetic field coincide each other approximately. However, we compare this situation with the case in which $J_m=0$, $H_c^\pm$ does not coincide with the saturation field (see section 6). The certain case $J-J_2=0$ is interesting, because we can see here magnetic entanglement with different values ($C(\rho)=1$ and $C(\rho)=1/2$) (figure (6)), while in the case of $J_m=0$, these two regimes cannot coexist for a fixed values of J and $J_2$ and only one sudden transition occurs. It is noticeable that, the behavior of quantum discord is similar to entanglement which measure by concurrence and again two sudden transition appear. However, we can observe here for quantum discord two transition take place with different values $C(\rho)=1$ and $C(\rho)=0.4$. Remarkably, GMQD has similar behavior to two above mentioned measures, only it show lesser value than the others. By comparison of quantum correlations in figures(7) and (8) which plotted versus magnetic field and temperature, we can see that again like as the previous section entanglement sudden death occurs while quantum discord and GMQD decrease asymptotically. Moreover, one can observe that by increasing the value of J and $J_2$ the peak of quantum correlations become wider.
 \begin{figure}
\includegraphics[width=3in]{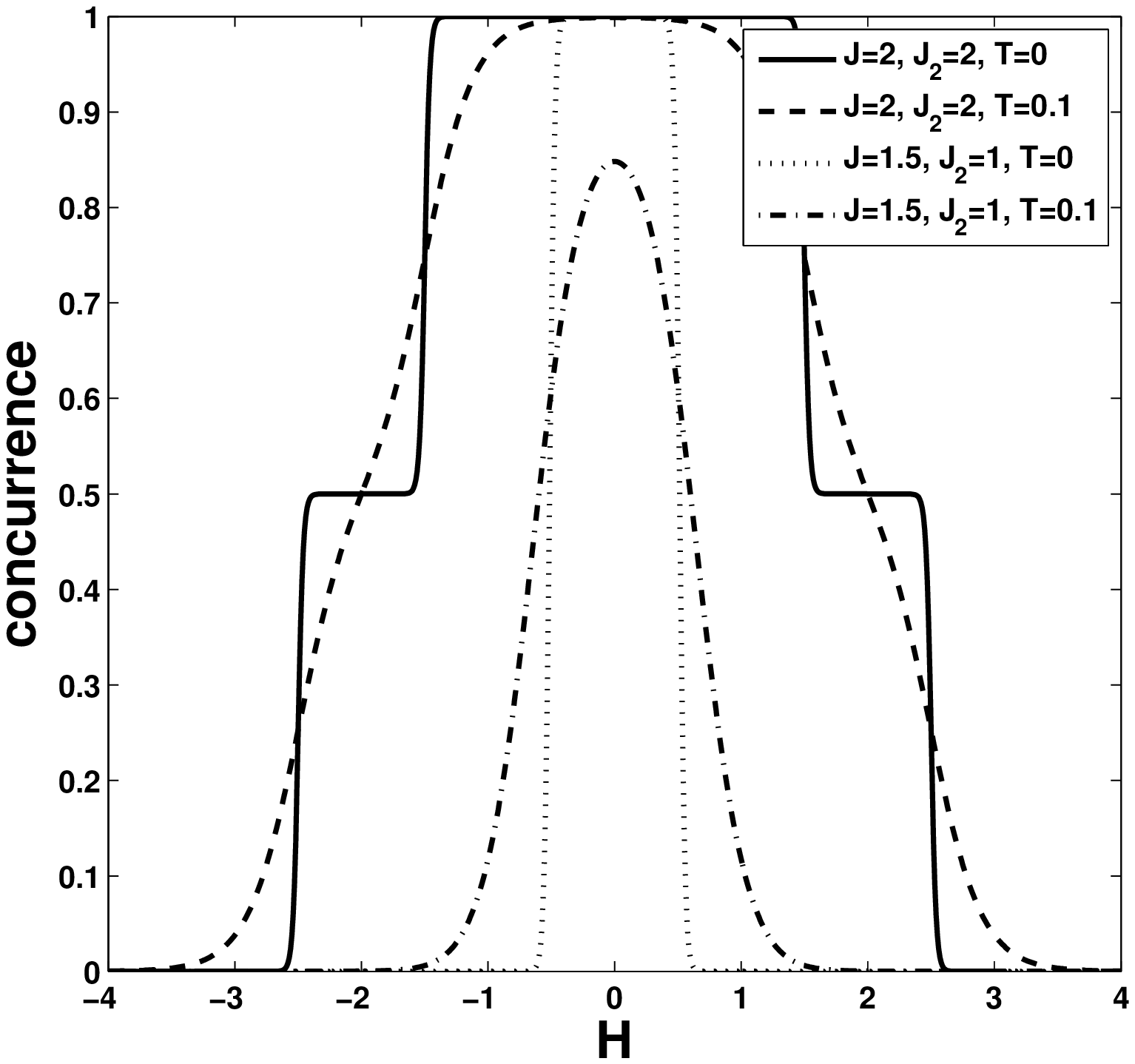}
\includegraphics[width=3in]{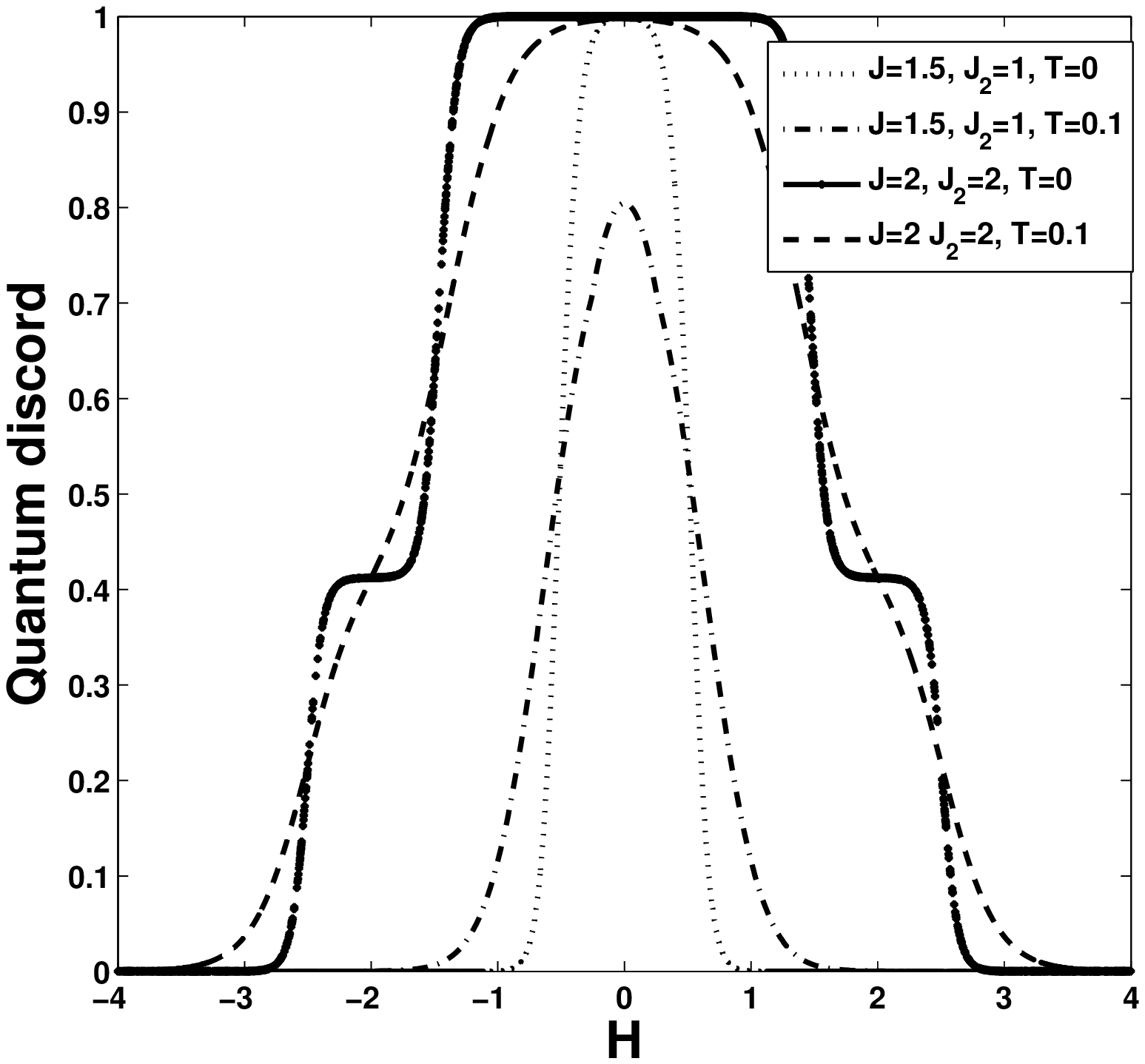}
\includegraphics[width=3in]{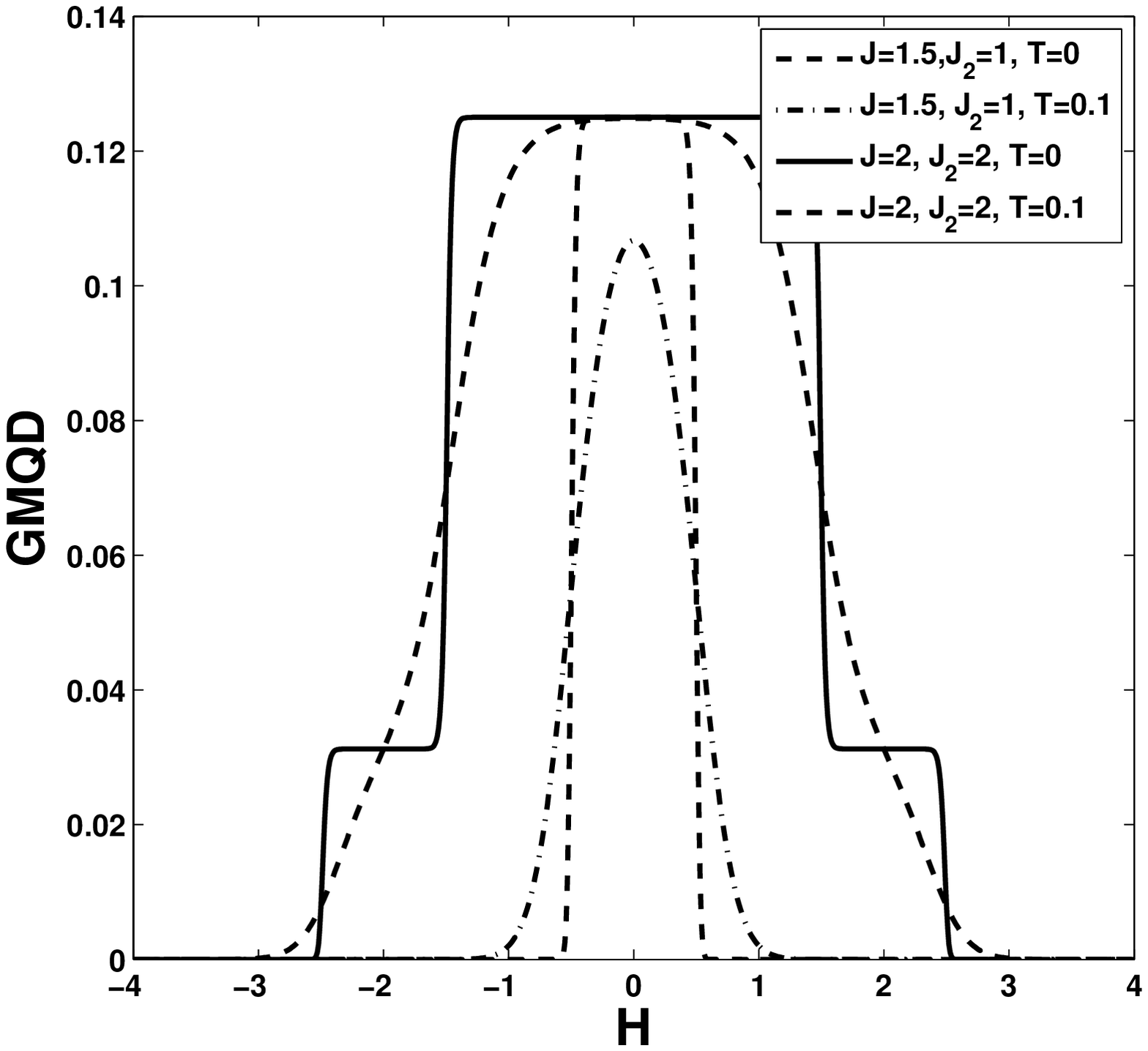}

\caption{Quantum correlations versus H for different values of temperature, $J_2$ and J and $J_m=1.5$. (a) Concurrence (b) QD (c) GMQD}
 \label{fig5}
\end{figure}

\begin{figure}
\includegraphics[width=3in]{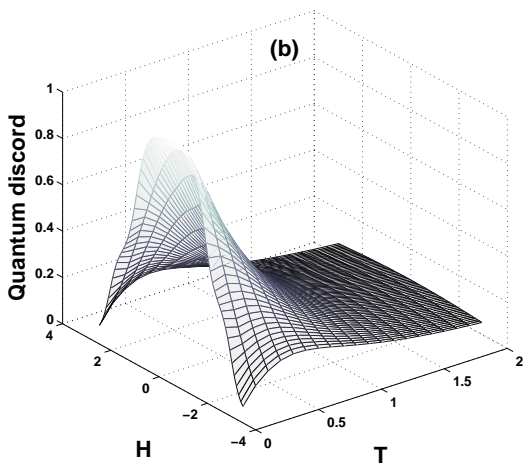}
\caption{Quantum discord versus magnetic field H and temperature T for$J_2=2$ and $J=2$ and $J_m=1.5$.}
 \label{fig6}
\end{figure}

\begin{figure}
\includegraphics[width=3in]{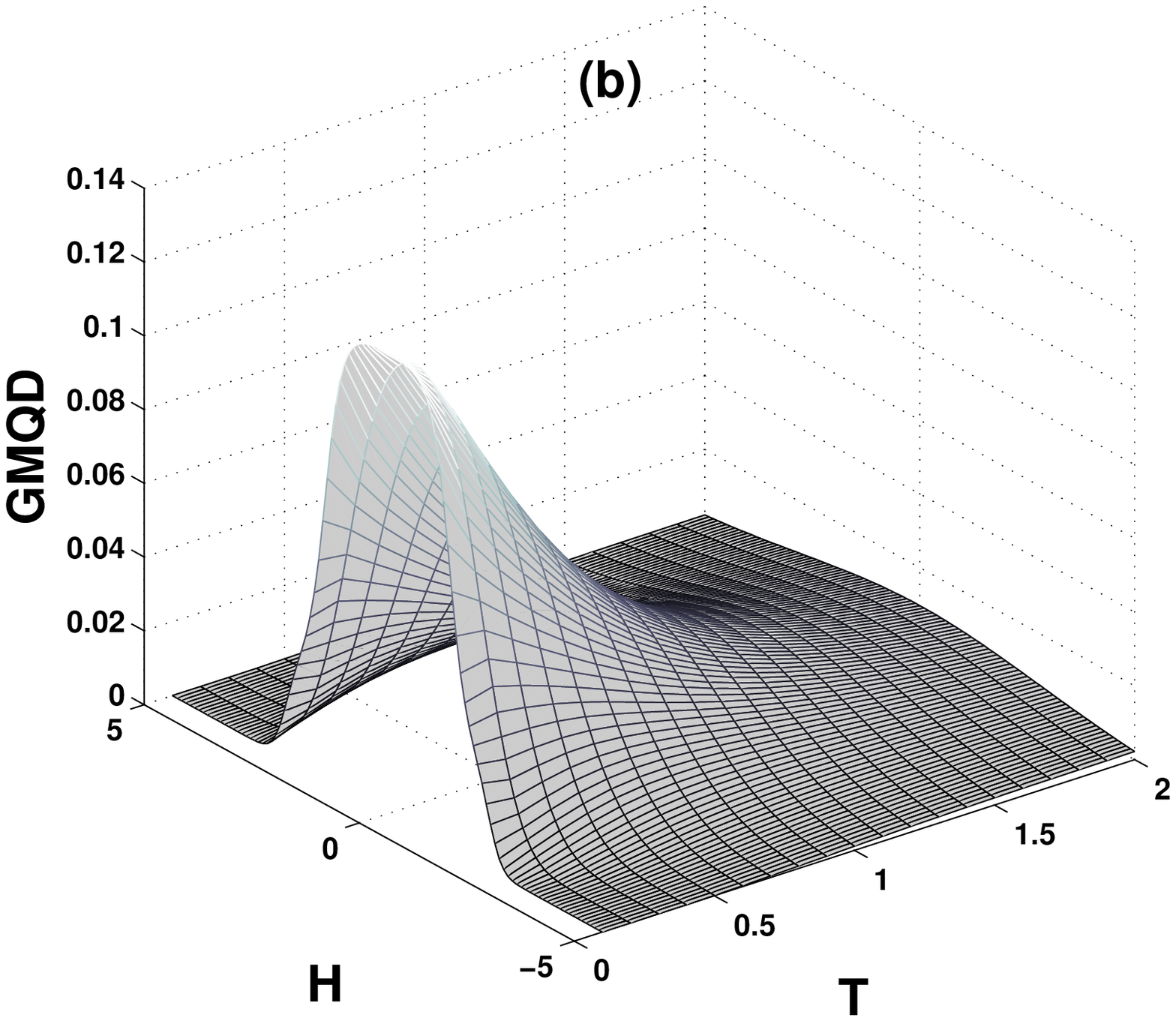}
\caption{GMQD versus magnetic field H and temperature T for $J_2=2$ and $J=2$ and $J_m=1.5$.}
 \label{fig7}
\end{figure}

\section{Conclusion}
We have studied the quantum correlations properties for a  S=1/2 Ising- Heisenberg model on a symmetrical diamond chain. We see that quantum correlations for an ideal diamond chain show symmetrical behavior respect to coupling constant in every temperature in both ferromagnetic and anti- ferromagnetic region.  Moreover, we observe quantum correlations have different order as a function of temperature T. We should mention that their order do not preserved for different coupling constant J. Therefore, no definite ordering relations between these quantum correlations exist. The observations are not in agreement with the previous information. As regard to the influence of temperature, one find that the higher the temperature is, the smaller quantum correlations are. Specifically, the concurrence contains finite threshold temperature, while the threshold temperature of QD and 1- norm is infinitely. We have also considered the effect of magnetic field and next nearest neighbor interaction ($J_m$) on a diamond chain. We observe that, the stronger the external magnetic field is, the smaller quantum correlation is. For the special case $J=J_2, J_m\neq0$ and T=0, we find that the quantum correlations experience two sudden transition when the absolute value of magnetic field increases in a range. This corresponds to the change of the ground state of the system. Additionally, the concurrence takes a zero value in a large range of the parameters and QD and 1- norm GQD takes a small value larger than zero. In this sense, we can say that the QD and 1- norm GQD are more robust than the concurrence.\\

\end{document}